\documentclass[12pt,a4paper]{article}

\usepackage[nosort]{cite}
\usepackage[hyperref,bulletsep]{collect}
\usepackage{graphicx}
\usepackage{pst-all}
\usepackage{bbm}
\usepackage{amsmath}
\usepackage{amssymb}
\usepackage{subfigure}

\makeatletter \@addtoreset{equation}{section} \makeatother

\makeatletter
\let\old@startsection=\@startsection
\let\oldl@section=\l@section
\renewcommand{\@startsection}[6]{\old@startsection{#1}{#2}{#3}{#4}{#5}{#6\mathversion{bold}}}
\renewcommand{\l@section}[2]{\oldl@section{\mathversion{bold}#1}{#2}}
\makeatother

\makeatletter
\let\old@makecaption=\@makecaption
\def\@makecaption{\small\old@makecaption}
\makeatother

\renewcommand{\thefootnote}{\arabic{footnote}}
\renewcommand{\leq}{\leqslant}

\let\oldPhi=\Phi
\let\oldPsi=\Psi
\let\oldGamma=\Gamma
\let\oldDelta=\Delta
\let\oldSigma=\Sigma
\let\oldTheta=\Theta
\let\oldPi=\Pi
\let\oldUpsilon=\Upsilon
\renewcommand{\Phi}{\mathnormal{\oldPhi}}
\renewcommand{\Psi}{\mathnormal{\oldPsi}}
\renewcommand{\Gamma}{\mathnormal{\oldGamma}}
\renewcommand{\Sigma}{\mathnormal{\oldSigma}}
\renewcommand{\Delta}{\mathnormal{\oldDelta}}
\renewcommand{\Theta}{\mathnormal{\oldTheta}}
\renewcommand{\Pi}{\mathnormal{\oldPi}}
\renewcommand{\Upsilon}{\mathnormal{\oldUpsilon}}

\newcommand{\NN}{\mathcal{N}}

\newcommand{\Lagr}{\mathcal{L}}

\newcommand{\tr}{\mathop{\mathrm{tr}}}

\newcommand{\OO}{\mathcal{O}}

\ifx\genfrac\sdflkaj

\else

\fi
\newcommand{\sfrac}[2]{{\textstyle\frac{#1}{#2}}}
\newcommand{\half}{\sfrac{1}{2}}

\newcommand{\bN}{\overline N}
\newcommand{\da}{\dot a}
\newcommand{\db}{\dot b}
\newcommand{\eps}{\epsilon}
\newcommand{\veps}{\varepsilon}
\newcommand{\la}{\lambda}

\newcommand{\bV}{\overline V}

\newcommand{\btau}{{\bar \tau}}
\newcommand{\ba}{{\bar a}}
\newcommand{\bb}{{\bar b}}

\newcommand{\bc}{{\bar c}}
\newcommand{\al}{\alpha}
\newcommand{\psl}{\makebox[0pt][l]{\hspace{1.2pt}$/$}p}
\newcommand{\qsl}{\makebox[0pt][l]{\hspace{.2pt}$/$}q}
\newcommand{\om}{\omega}
\newcommand{\s}{\sigma}

\newcommand{\odd}{\mbox{\tiny odd}}
\newcommand{\even}{\mbox{\tiny even}}

\newcommand{\nn}{\nonumber}

\def\[{\begin{equation}}
\def\]{\end{equation}}

\makeatletter
\def\mr@ignsp#1 {\ifx\:#1\@empty\else #1\expandafter\mr@ignsp\fi}%
\newcommand{\multiref}[1]{\begingroup
\xdef\mr@no@sparg{\expandafter\mr@ignsp#1 \: }%
\def\mr@comma{}%
\@for\mr@refs:=\mr@no@sparg\do{\mr@comma\def\mr@comma{,}\ref{\mr@refs}}%
\endgroup}
\makeatother

\newcommand{\hypref}[2]{\ifx\href\asklfhas #2\else\href{#1}{#2}\fi}

\renewcommand{\eqref}[1]{(\multiref{#1})}

\ifx\href\asklfhas\newcommand{\href}[2]{#2}\fi

\newcommand{\be}{\begin{eqnarray}}
\newcommand{\ee}{\end{eqnarray}}

\begin{document}


\thispagestyle{empty}
\begin{flushright}\footnotesize
\texttt{ITEP-TH-30/08}\\ \texttt{LPTENS-08/32}\\
\texttt{UUITP-13/08}
\end{flushright}
\vspace{0.8cm}

\renewcommand{\thefootnote}{\fnsymbol{footnote}}
\setcounter{footnote}{0}

\begin{center}
{\Large\textbf{\mathversion{bold} The Bethe ansatz for
superconformal Chern-Simons }\par}

\vspace{1.5cm}

\textrm{J.~A.~Minahan$^{1}$ and K.~Zarembo$^{2,1}$\footnote{Also at
ITEP, Moscow, Russia}} \vspace{8mm}

\textit{$^{1}$ Department of Physics and Astronomy, Uppsala University\\
SE-751 08 Uppsala, Sweden}\\
\texttt{joseph.minahan, konstantin.zarembo@fysast.uu.se}
\vspace{3mm}

\textit{$^{2}$ Laboratoire de Physique Th\'eorique de l'Ecole
Normale Sup\'erieure\\ 24 rue Lhomond, Paris CEDEX 75231, France}
\vspace{3mm}


\par\vspace{1cm}

\textbf{Abstract} \vspace{5mm}

\begin{minipage}{14cm}
We study the anomalous dimensions for scalar operators for a
three-dimensional Chern-Simons theory recently proposed in
arXiv:0806.1218. We show that the mixing matrix at two-loop order is
that for an integrable Hamiltonian of  an $SU(4)$ spin chain with
sites alternating between the fundamental and the anti-fundamental
representations. We find a set of Bethe equations from which the
anomalous dimensions can be determined and give a proposal for the
Bethe equations to the full superconformal group of $OSp(2,2|6)$.

\end{minipage}

\end{center}

\vspace{0.5cm}

\newpage
\setcounter{page}{1}
\renewcommand{\thefootnote}{\arabic{footnote}}
\setcounter{footnote}{0}

\section{Introduction}

Integrability has proven to be a powerful tool in analyzing $\NN=4$
Super Yang-Mills in the planar limit.   An interesting question is
whether or not there are other gauge theories with a high degree of
supersymmetry that are also integrable at the planar level.

Recently, a proposal was made by Aharony, Bergman, Jafferis and
Mal\-da\-ce\-na \linebreak(ABJM) \cite{Aharony:2008ug}, following a
large body of work on multiple M2-branes
\cite{Schwarz:2004yj,Bagger:2006sk,Gustavsson:2007vu,Bagger:2007jr,Bagger:2007vi,VanRaamsdonk:2008ft,Distler:2008mk,Ho:2008ei,Gomis:2008be},
for a three dimensional superconformal $SU(N)\times SU(N)$
Chern-Simons  theory that seems to be the effective theory for a
stack of M2 branes at a $Z_k$ orbifold point. In the large $N$
limit, the gravitational dual becomes M-theory on $AdS_4\times
S^7/Z_k$.  The integer $k$ is the level of the first $SU(N)$ and the
level of the second $SU(N)$ is $-k$. The theory has manifest
$SU(2)\times SU(2)\times U(1)$ $R$-symmetry and two sets of scalar
fields transforming in bifundamental representations of $SU(N)\times
SU(N)$.  The first set of scalars, $A_a$ are doublets under one
$SU(2)$ of the $R$-symmetry group and transform in the $(N,\bN)$
representation and the second set of scalars $B_{\da}$ are doublets
under the second $SU(2)$ and transform under the $
(\bN,N)$ representation.

The scalars can be conveniently expressed as $N\times N$ matrices,
in which case the superpotential takes the form \be
W=\frac{2\pi}{k}\eps^{ab}\eps^{\da\db}\tr(A_aB_{\da} A_b B_{\db})\,.
\ee Remarkably, as was argued in \cite{Aharony:2008ug} and proven in
\cite{Benna:2008zy},  the $R$-symmetry is enhanced to $SO(6)$ due to
contributions from the Chern-Simons terms, and the theory has
$\NN$=6 supersymmetry if $k>2$.  If $k=1$ or $2$, then there is
$\NN=8$ supersymmetry.

The ABJM model has the large-$N$ limit with the 't Hooft coupling
$\la=N/k$ \cite{Aharony:2008ug}. For infinite $N$ and finite $\la$,
$k$ is infinite and $\la$ is essentially continuous. In the case of
large $k$, the orbifold effectively compactifies to a cylinder and
M-theory approaches type IIA string theory on $AdS_4\times CP^3$.
String theory propagating on this space is classically integrable,
so one might expect integrability to appear in the dual gauge theory
as well. The classical limit of string theory corresponds to
$\lambda \gg 1$. We will analyze the opposite regime of weak
coupling.

The scalar fields can be grouped into $SU(4)$ multiplets $Y^A$ as
follows \be Y^A=(A_1,A_2,B^\dag_{\dot1},B^\dag_{\dot2})\qquad
Y^\dag_A=(A^\dag_1,A^\dag_2,B_{\dot1},B_{\dot 2})\,, \ee and a class
of gauge invariant operators can be built out of these scalars in
the form\footnote{Recently considered BMN operators in the ABJM
model \cite{Nishioka:2008gz} are particular cases of these more
general operators, which as a matter of fact resemble scalar
operators in the orbifolds of $\mathcal{N}=4$ super-Yang-Mills in
four dimensions \cite{Astolfi:2006is}. } \be\label{ops}
\OO=\tr(Y^{A_1}Y^\dag_{B_1}Y^{A_2}Y^\dag_{B_2}\dots
Y^{A_L}Y^\dag_{B_L})\chi^{B_1\dots B_L}_{A_1\dots A_L}\,. \ee The
bare dimension of $\OO$ is $L$ and $\OO$ is a chiral primary if
$\chi$ is symmetric in all $A_i$ indices, symmetric in all $B_i$
indices and all traces are zero.  If $\OO$ is not a chiral primary,
then it has a nonzero anomalous dimension.   The leading order
contribution to the anomalous dimension comes at two-loop order
since the contributions all come with even powers of $k$,  and in
general leads to operator mixing.

In this paper we compute the leading order operator mixing matrix
for the scalar operators in (\ref{ops}) and show that it is
isomorphic to an integrable Hamiltonian of an $SU(4)$ spin-chain
with sites alternating between the fundamental and anti-fundamental
representations and next to nearest neighbor interactions.  The
details of the calculation parallel the arguments in
\cite{Minahan:2002ve} for scalar operators in $\NN=4$ SYM${}_4$. In
that case there was also $SU(4)$ $R$-symmetry and the scalar
operators had a mixing matrix that is isomorphic to a Hamiltonian
for an integrable $SU(4)$ spin chain.  One can then find a set of
Bethe equations whose solutions lead to the eigenvalues of the
mixing matrix.  For the ABJM model we will also find a set of Bethe
equations, but with different weights than the $\NN=4$ SYM${}_4$
case.  One can then try extending the calculation to the full
superconformal group as in \cite{Beisert:2003jj,Beisert:2003yb}.
While we do not compute the Hamiltonian explicitly, we propose a
natural extension of the $SU(4)$ chain to an $OSp(2,2|6)$ chain.
Extension to higher orders in $\lambda $ should be also possible as
in the $\mathcal{N}=4$, $D=4$ super-Yang-Mills
\cite{Beisert:2003tq,Beisert:2005fw,Beisert:2006ez,Beisert:2006ib},
but will not be discussed in this paper.

Previously, Gaiotto and Yin \cite{Gaiotto:2007qi}
studied   a different version of a supersymmetric Chern-Simons
theory, with lower supersymmetry and an $SU(2)$ $R$-symmetry group.
In that theory the gauge invariant operators can be mapped to an
$SU(2)$ spin chain with both nearest neighbor and next to nearest
neighbor interactions, and so the theory cannot be integrable.  In
the ABJM theory the larger R-symmetry group and cancelation of the
nearest neighbor interactions allow integrability to be possible.

In section 2 we construct the Hamiltonian by explicitly computing a
two-loop Feynman diagram containing a six-point vertex, a
two-loop diagram containing a fermion loop and a two loop diagram containing gauge propagators.  The resulting
Hamiltonian has next to nearest neighbor interactions of two types.
It turns out that the nearest neighbor interactions cancel out between the three types of diagrams.  In
section 3 we show that the resulting Hamiltonian is integrable and
we find the corresponding Bethe equations for this system.  In
section 4 we propose an extension of the Bethe equations to the full
$OSp(2,2|6)$ superconformal group.  In section 5 we summarize our
results and offer suggestions for further study.  In appendices we
give some technical details and consider the explicit example of an
operator with four sites.


\section{Two-loop amplitudes and the Hamiltonian}
The operators (\ref{ops}) need to be renormalized to make their
correlation functions finite. Transition to the basis where the
renormalization is multiplicative leads to the operator mixing:
\begin{equation}\label{}
\mathcal{O}^{\bf A}_{\rm ren}=Z^{\bf A}{}_{\bf B}(\Lambda )
\mathcal{O}^{\bf B}_{\rm bare},
\end{equation}
where ${\bf A}$ is a multi-index that enumerates all possible
operators, $\Lambda $ is a UV cutoff, and the $Z$-factor subtracts
all the UV divergences from the correlation functions. The mixing
matrix (the quantum part of the dilatation operator) is defined as
\begin{equation}\label{}
\Gamma =Z^{-1}\,\frac{dZ}{d\ln\Lambda }\,.
\end{equation}
Its eigenstates are conformal operators and the eigenvalues are
their anomalous dimensions.

It convenient to represent the operators (\ref{ops}) as states in a
quantum spin chain with $2L$ sites.
\begin{figure}[t]
\centerline{\includegraphics[width=8cm]{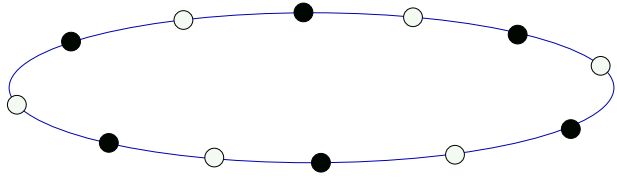}}
\caption{\label{altr}\small The alternating spin chain.}
\end{figure}
The spin is alternating between the
fundamental representation of $su(4)$ on odd sites and the
anti-fundamental representation on the even sites (fig.~\ref{altr}).
The mixing matrix can then be regarded as the Hamiltonian acting in
the Hilbert space $(V\otimes\bar{V})^{\otimes L}$, where $V$
($\bar{V}$) is the the $\mathbf{4}$ ($\bar{\mathbf{4}}$) of $SU(4)$.
We will compute this Hamiltonian to the lowest order in $\lambda $
and in $1/N$.

The action of the $\mathcal{N}=6$ Chern-Simons \cite{Aharony:2008ug}
is \footnote{The fermion terms in the action \cite{Benna:2008zy} are
listed in the appendix~\ref{compferm}.}
\begin{eqnarray}\label{}
S&=&\frac{k}{4\pi }\int_{}^{}d^3x\,\,{\rm tr}\,\left[
\varepsilon ^{\mu \nu \lambda }\left(
A_\mu \partial _\nu A_\lambda +\frac{2}{3}\,A_\mu A_\nu A_\lambda
-\hat{A}_\mu \partial _\nu \hat{A}_\lambda
-\frac{2}{3}\,\hat{A}_\mu \hat{A}_\nu \hat{A}_\lambda
\right)+  D_\mu Y^\dagger _AD^\mu Y^A
\right.
\nonumber \\
&&\left.+\frac{1}{12}\,Y^AY^\dagger _AY^BY^\dagger _BY^CY^\dagger _C
+\frac{1}{12}\,Y^AY^\dagger _BY^BY^\dagger _CY^CY^\dagger _A
-\frac{1}{2}\,Y^AY^\dagger _AY^BY^\dagger _CY^CY^\dagger _B
\right.
\nonumber \\
&&\left.
+\frac{1}{3}\,Y^AY^\dagger _BY^CY^\dagger _AY^BY^\dagger _C
+{\rm fermions}\right],
\end{eqnarray}
where $D_\mu Y=\partial Y+A_\mu Y-Y\hat{A}_\mu $, $D_\mu Y^\dagger
=\partial _\mu Y^\dagger +\hat{A}_\mu Y^\dagger -Y^\dagger A_\mu $.
\begin{figure}[t]
\centerline{\includegraphics[totalheight=0.12\textheight]{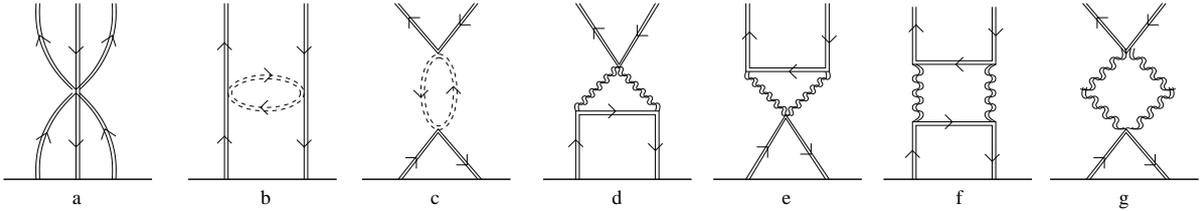}}
\caption{\label{diagrams}\small The planar diagrams that contribute
to operator mixing  at two loops. The horizontal bar denotes the
operator. The directions of the arrows refer to the flow of the
$SU(4)$ flavor. Since the superpartners of the scalars are in the
conjugate representation of $SU(4)$, the fermion arrows in (b) and (c)
have the opposite orientation.  The gauge propagators in (d), (e), (f) and (g) do not have arrows since they do not carry $SU(4)$ charges. It turns out that only (a), (b) and (d) contribute to the anomalous dimension.}
\end{figure}
Since the interactions are of the $Y^6$
type (and $Y^2\Psi^2 $, if we include fermions), the lowest order
contribution to the mixing matrix arises at two loops
(fig.~\ref{diagrams}). The scalar diagram (a) connects three sites
on the spin chain, so the Hamiltonian will involve interactions of
three adjacent spins:
\begin{equation}\label{}
\Gamma=\frac{\lambda ^2}{4}\sum_{l=1}^{2L}H_{l,l+1,l+2},
\end{equation}
where $H_{l,l+1,l+2}$ acts on $\bar{V}\otimes V\otimes\bar{V}$ for
$l$ even and on $V\otimes \bar{V}\otimes V$ for $l$ odd.
The diagrams in (b) and (c) contribute
only to the nearest-neighbor interactions. And finally there are
also diagrams with the gauge-boson exchange and the self-energy
graphs. We will compute the scalar diagram here and the other diagrams
in appendix~\ref{compferm}. The gauge-boson exchange and
self-energy contribute only to the diagonal term in the Hamiltonian,
and we will reconstruct them by supersymmetry.

The loop integral in the scalar diagram can be easily calculated in
the coordinate representation:
$$
\int_{}^{}d^3x\,\left(\frac{1}{4\pi |x|}\right)^3=\frac{1}{16\pi
^2}\,\ln\Lambda\,.
$$
This contains three 3-dimensional propagators in the coordinate
representation. The non-trivial part is combinatorics of the $SU(4)$
indices, which can be handled graphically.
\begin{figure}[t]
\centerline{\includegraphics[width=12cm]{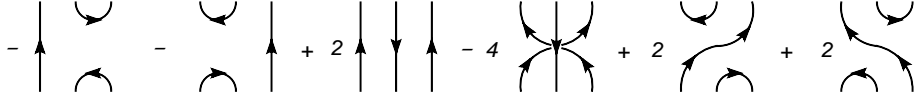}}
\caption{\label{ham}\small The two-loop Hamiltonian. The arrows
denote $SU(4)$ index contractions.}
\end{figure}
Omitting unnecessary details, we show the odd-site Hamiltonian in
fig.~\ref{ham}. The even-site Hamiltonian is obtained by flipping
the arrows.
\begin{figure}[t]
\centerline{\includegraphics[width=6cm]{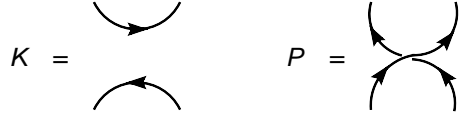}}
\caption{\label{kandp}\small The permutation and trace operators.}
\end{figure}
The Hamiltonian can be expressed in terms
of the two basic operators (fig.~\ref{kandp}): the permutation
$P:V\otimes V\rightarrow V\otimes V$ (or
$P:\bar{V}\otimes\bar{V}\rightarrow \bar{V}\otimes\bar{V}$) and the
trace $K:V\otimes\bar{V}\rightarrow V\otimes\bar{V}$ (or
$K:\bar{V}\otimes V\rightarrow \bar{V}\otimes V$), defined as
\begin{eqnarray}\label{PKdef}
P^{AB}{}_{A'B'}&=&\delta ^A{}_{B'}\,\delta ^B{}_{A'}\nonumber \\
K^A_B\,{}^{B'}_{A'}&=&\delta ^A{}_{A'}\,\,\delta _B{}^{B'}.
\end{eqnarray}
The spin-chain operator in fig.~\ref{ham} then reads
\begin{equation}\label{hscalar}
\Gamma _{\rm sc}=\frac{\lambda ^2}{2}\sum_{l=1}^{2L}\left(
-K_{l,l+1}+1-2P_{l,l+2}+P_{l,l+2}K_{l,l+1}+K_{l,l+1}P_{l,l+2}
\right)
\end{equation}

If we add the fermion loops and gauge contributions from
appendix~\ref{compferm}, a remarkable cancelation happens. These
terms contribute the two-site trace operator with the coefficient
$+\lambda ^2/2$, which exactly cancels the first term in
(\ref{hscalar}) and leaves no  nearest-neighbor terms in the
Hamiltonian. We have not computed  the constant piece, but supersymmetry requires that the
ground state energy is zero, which happens when the constant term
and the permutation combine into a projection $1-P$ on the symmetric
traceless states. {}From this we can find the missing constant
contribution and get the full two-loop dilatation operator:
\begin{equation}\label{dilop}
\Gamma =\frac{\lambda ^2}{2}\sum_{l=1}^{2L}\left(
2-2P_{l,l+2}+P_{l,l+2}K_{l,l+1}+K_{l,l+1}P_{l,l+2}
\right).
\end{equation}
This is our main result.

The ground states of the Hamiltonian are symmetric traceless chiral
primary operators. Because chiral primaries are protected by
supersymmetry, it should be possible to directly compare their
spectrum with the supergravity harmonics on $AdS_4\times CP^3$. We
have just checked that the chiral primaries are in one-to-one
correspondence with the spherical functions on $CP^3$
(appendix~\ref{appspfun}). We also worked out the complete spectrum
of the Hamiltonian for operators of length four ($L=2$) in
appendix~\ref{dim2}.

Is the Hamiltonian (\ref{dilop}) integrable? Integrable alternating
spin chains have been studied before
\cite{Faddeev:1985qu,Destri:1987ug,deVega:1991rc,abadrios,Martins99,Ribeiro:2005kn},
and although we were unable to find the Hamiltonian (\ref{dilop}) in
the literature, there is a general formalism \cite{deVega:1991rc}
that allows one to build an alternating integrable Hamiltonian in
any representations starting with appropriate R-matrices. The
resulting Hamiltonian indeed involves nearest-neighbor and
three-site interactions, but in general breaks charge conjugation
symmetry $\mathbf{4}\leftrightarrow\bar{\mathbf{4}}$. It turns out
for $SL(n)$ groups the nearest neighbor interactions always cancel
out. If one further makes a special choice of paramters then the
conjugation symmetry is  preserved and the spin-chain Hamiltonian
exactly coincides with the dilatation operator (\ref{dilop})! We can
then make use of the general formulas
\cite{Kulish:1983rd,Arnaudon:2004vd} that describe the spectrum via
the algebraic Bethe ansatz \cite{Faddeev:1996iy}.

\section{Integrability for an $SU(4)$ spin chain with alternating sites}

In this section we show that the Hamiltonian derived in the previous
section is that of an integrable $SU(4)$ spin chain with sites
alternating between the fundamental and anti-fundamental
representation.  We will actually generalize the derivation for any
$SU(n)$ group, specializing to $SU(4)$ at the end.

In order to establish integrability, one first defines an $R$-matrix
$R_{ab}(u)$ which is a linear map from a tensor product of two
vector spaces in the fundamental representation of $SU(n)$
\be\label{rmap} R_{ab}(u):\, V_a\otimes V_b\,\to\, V_a\otimes V_b\,,
\ee where the parameter $u$ is the spectral parameter.  If we let
\be\label{rmat} R_{ab}(u)=u-P_{ab}\,, \ee then $R_{ab}(u)$ satisfies
the Yang-Baxter equation \be\label{ybe}
R_{ab}(u-v)R_{ac}(u)R_{bc}(v)=R_{bc}(v)R_{ac}(u)R_{ab}(u-v)\,. \ee

The results in (\ref{rmap}), (\ref{rmat}) and (\ref{ybe}) can be
generalized to all representations using a universal $R$ matrix, but
for our purposes we only need the  cases where $V_1$ and $V_2$ are
the fundamental or anti-fundamental representations.  We therefore
introduce two other $R$-matrices \be
R_{a\bb}(u)&=&u+K_{a\bb}\nn\\
R_{\ba\bb}(u)&=&u-P_{\ba\bb} \ee where $P_{\ba\bb}$ and $K_{a\bb}$
were defined in (\ref{PKdef}). We then have the additional
Yang-Baxter equations 
\be\label{ybeb}
R_{\ba\bb}(u-v)R_{\ba c}(u)R_{\bb c}(v)&=&R_{\bb c}(v)R_{\ba c}(u)R_{\ba\bb}(u-v)\nn\\
R_{ab}(u-v)R_{a \bc}(u)R_{b\bc}(v)&=&R_{b \bc}(v)R_{a
\bc}(u)R_{ab}(u-v). \ee To show these formulae, the following
identities are useful:
$$ P_{ab}P_{ab}=1\qquad
K_{a\bb}K_{a\bb}=nK_{a\bb}\qquad P_{ab}K_{b\bc}=K_{a\bc}K_{b\bc}\,.
$$
In addition, there are a set of modified Yang-Baxter equations
\be\label{ybem}
R_{a\bb}(u-v-n)R_{a c}(u)R_{\bb c}(v)&=&R_{\bb c}(v)R_{a c}(u)R_{a\bb}(u-v-n)\nn\\
R_{\ba b}(u-v-n)R_{\ba \bc}(u)R_{b\bc}(v)&=&R_{b \bc}(v)R_{\ba
\bc}(u)R_{\ba b}(u-v-n) \ee

Given these $R$-matrices we can construct the  monodromy matrices
$T_a(u,\al)$ and $T_{\ba}(u,\al)$ \be
T_a(u,\al)&=&C\, R_{a1}(u)R_{a\bar 1}(u+\al)R_{a2}(u)R_{a\bar 2}(u+\al)\dots R_{aL}(u)R_{a\bar L}(u+\al)\nn\\
T_{\ba}(u,\al)&=&C\, R_{\ba1}(u+\al)R_{\ba\bar
1}(u)R_{\ba2}(u+\al)R_{\ba\bar 2}(u)\dots R_{\ba L}(u+\al)R_{\ba\bar
L}(u)\nn \ee where $a$ and ${\ba}$ refer to auxiliary spaces in the
fundamental and anti-fundamental representations, $\alpha$ is a
constant parameter and $C$ is a normalization constant.  It then
follows from the Yang-Baxter equations in (\ref{ybe}) and (\ref{ybeb}) that
\be\label{ybeT}
R_{ab}(u-v)T_a(u,\al)T_b(v,\al)&=&T_b(v,\al)T_a(u,\al)R_{ab}(u-v)\nn\\
R_{\ba\bb}(u-v)T_\ba(u,\al)T_\bb(v,\al)&=&T_\bb(v,\al)T_\ba(u,\al)R_{\ba\bb}(u-v)
\ee where $b$ ($\bb$) refers to a different fundamental
(anti-fundamental) auxiliary space, but otherwise $T_b(v,\al)$ and
$T_\bb(v,\al)$ act on the same $(V\otimes\bV)^L$ space. If we define
the transfer matrices $\tau(u)$ and $\btau(u)$ as the trace of
$T_a(u,\alpha)$ and $T_\bb(u,\alpha)$ over the auxiliary spaces, \be
\tau(u,\alpha)={\tr}_a T_a(u,\al) \qquad\btau(u,\alpha)={\tr}_\ba
T_\ba(u,\alpha) \ee then (\ref{ybeT}) leads to 
\be
[\tau(u,\alpha),\tau(v,\alpha)]=0\qquad
[\bar{\tau}(u,\alpha),\bar{\tau}(v,\alpha)]=0 \ee for any $u$ and $v$.  Since
$\tau(u,\alpha)$ and $\btau(u,\alpha)$ are  polynomials of order
$2L$, each one gives up to $2L$ independent commuting quantities. Of
particular interest are $\tau(0,\alpha)$ and $\btau(0,\alpha)$ \be
\tau(0,\alpha)&=&C\left(\frac{1}{\al(n+\al)}\right)^L\prod_{i=1}^L(\al+K_{2i-1,2i})\prod P_{2L-2i+1,2L-2i-1}\nn\\
\btau(0,\alpha)&=&C\left(\frac{1}{\al(n+\al)}\right)^L\prod_{i=1}^L(\al+K_{2i,2i+1})\prod
P_{2L-2i+2,2L-2i}\nn \ee which are the analogs of the shift operator
for a homogeneous chain, and the two Hamiltonians \be
H_{\odd}&=&(\tau(0,\alpha))^{-1}\frac{d}{du}\tau(u,\alpha)\Big|_{u=0}\nn\\
&=&\sum_{i=1}^L\left(\frac{1}{\al}-P_{2i-1,2i+1}-\frac{1}{\al}K_{2i-1,2i}K_{2i,2i+1}+\frac{1}{n+\al}K_{2i,2i+1}K_{2i-1,2i}\right)\,\nn\\
H_{\even}&=&(\tau(0,\alpha))^{-1}\frac{d}{du}\tau(u,\alpha)\Big|_{u=0}\nn\\
&=&\sum_{i=1}^L\left(\frac{1}{\al}-P_{2i,2i+2}-\frac{1}{\al}K_{2i,2i+1}K_{2i,2i+2}+\frac{1}{n+\al}K_{2i+1,2i+2}K_{2i,2i+1}\right)\,\nn\,.
\ee

We can see that $H_{\odd}$ and $H_{\even}$ are proportional to the
contribution of the odd and even sites in the gauge theory spin
chain if $n=4$ and $\al=-n/2$.  However, we still have to establish
that $[\tau(u,\alpha),\btau(v,\alpha)]=0$ in order that
$H_{\odd}+H_{\even}$ is an integrable Hamiltonian. In order to show
this there should be a Yang-Baxter equation of the form
\be\label{ybep}
R_{a\bb}(u-v+\beta)T_a(u,\alpha)T_{\bb}(v,\alpha)=T_{\bb}(v,\alpha)T_a(u,\alpha)R_{a\bb}(u-v+\beta)\,,
\ee where $\beta$ can be any constant. In order for (\ref{ybep}) to
work, we have to satisfy {\it both} the equations \be\label{constr}
R_{a\bb}(u-v+\beta)R_{ac}(u)R_{\bb c}(v+\al)&=&R_{\bb c}(v+\al)R_{ac}(u)R_{a\bb}(u-v+\beta)\nn\\
R_{a\bb}(u-v+\beta)R_{a \bar c}(u+\al)R_{\bar b \bar c}(v)&=&R_{\bar
b \bar c}(v)R_{a \bar c}(u+\al)R_{a\bb}(u-v+\beta)\,. \ee Using
(\ref{ybeb}) and (\ref{ybem}), we see that both equations in
(\ref{constr}) are true only if $\beta=\al=-n/2$.  But this is
precisely the value of $\al$ that matches the Hamiltonian derived
from the gauge theory spin chain!  Furthermore, if $\al=-n/2$ and we
now choose $C=(2/n)^L$, then the product of $\tau(0,-n/2)$ and
$\btau(0,-n/2)$ is
\be
\tau(0,-n/2)\btau(0,-n/2)=\prod_{i=1}^{2L}P_{2L+2-i,2L-i}\,, \ee
which is the operator that shifts every flavor index by two sites.
Since the trace is invariant under such a shift,  we must have
$\tau(0,-n/2)\btau(0,-n/2)\left|{\rm phys}\right\rangle=\left|{\rm
phys}\right\rangle$ for all physical operators.

{}From now on we let $\al=-n/2$ and define
$\tau(u)\equiv\tau(u,-n/2)$, $\btau(u)\equiv\btau(u,-n/2)$.  One can
construct the eigenvalues for $\tau(u)$ and $\btau(u)$ using the
algebraic Bethe ansatz.  This was originally done in \cite{Kulish:1983rd} for an
inhomogenous spin chain with different representations on the $2L$
sites.  For the case considered here, one finds the eigenvalues of
$\Lambda(u)$ are
\begin{eqnarray}\label{traneig}
\Lambda (u)&=&(u-1)^L(u-2)^L\prod_{j=1}^{M_u}\frac{u-iu_j+\frac{1}{2}}{u-iu_j-\frac{1}{2}}
+u^L(u-2)^L\prod_{j=1}^{M_u}\frac{u-iu_j-\frac{3}{2}}{u-iu_j-\frac{1}{2}}\,
\prod_{k=1}^{M_r}\frac{u-ir_k}{u-ir_k-1}
\nonumber \\
&&+u^L(u-2)^L\prod_{n=1}^{M_v}\frac{u-iv_n-\frac{1}{2}}{u-iv_n-\frac{3}{2}}\,
\prod_{k=1}^{M_r}\frac{u-ir_k-2}{u-ir_k-1}
+u^L(u-1)^L\prod_{n=1}^{M_v}\frac{u-iv_n-\frac{5}{2}}{u-iv_n-\frac{3}{2}}\,.\nn\\
\end{eqnarray}
where the $u_j$, $v_j$ and $r_j$ are a set of  Bethe roots
associated with the $SU(4)$ Dynkin diagram shown in figure 2.
\begin{figure}[t]
\centerline{\includegraphics[totalheight=0.1\textheight]{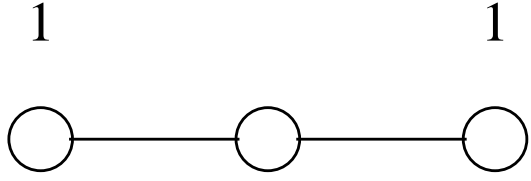}}
\caption{\label{su4dyn}\small The $SU(4)$ Dynkin diagram where the
numbers indicate the Dynkin labels of the representation.  The roots
$u_j$ are associated with one outer root, $v_j$ with the other outer
root, and $r_j$ with the middle root.}
\end{figure}
Since $\Lambda(u)$ is clearly a polynomial in $u$, the Bethe roots
must satisfy a set of Bethe equations to cancel off the poles in
(\ref{traneig}), \be\label{betheeq}
\left(\frac{u_j+i/2}{u_j-i/2}\right)^L&=&\prod_{k=1,k\ne j}^{M_u}\frac{u_j-u_k+i}{u_j-u_k-i} \prod_{k=1}^{M_r}\frac{u_j-r_k-i/2}{u_j-r_k+i/2}\nn\\
1&=& \prod_{k=1,k\ne j}^{M_r}\frac{r_j-r_k+i}{r_j-r_k-i}\prod_{k=1}^{M_u}\frac{r_j-u_k-i/2}{r_j-u_k+i/2}\prod_{k=1}^{M_v}\frac{r_j-v_k-i/2}{r_j-v_k+i/2}\nn\\
\left(\frac{v_j+i/2}{v_j-i/2}\right)^L&=&\prod_{k=1,k\ne
j}^{M_v}\frac{v_j-v_k+i}{v_j-v_k-i}
\prod_{k=1}^{M_r}\frac{v_j-r_k-i/2}{v_j-r_k+i/2}\,. \ee

The eigenvalues of $\bar{t}(u)$ can be found from the conjugation
condition  $\bar{\Lambda }(u)=\Lambda (2-u^*)^*$. We
find\footnote{Here we also use the fact that the Bethe roots are
real or come in the complex conjugate pairs.}:
\begin{eqnarray}\label{Lconj}
\bar{\Lambda} (u^*)&=&u^L(u-1)^L\prod_{j=1}^{M_u}\frac{u-iu_j-\frac{5}{2}}{u-iu_j-\frac{3}{2}}
+u^L(u-2)^L\prod_{j=1}^{M_u}\frac{u-iu_j-\frac{1}{2}}{u-iu_j-\frac{3}{2}}\,
\prod_{k=1}^{M_r}\frac{u-ir_k-2}{u-ir_k-1}
\nonumber \\
&&+u^L(u-2)^L\prod_{n=1}^{M_v}\frac{u-iv_n-\frac{3}{2}}{u-iv_n-\frac{1}{2}}\,
\prod_{k=1}^{M_r}\frac{u-ir_k}{u-ir_k-1}
+(u-1)^L(u-2)^L\prod_{n=1}^{M_v}\frac{u-iv_n+\frac{1}{2}}{u-iv_n-\frac{1}{2}}\,.\nn\\
\end{eqnarray}
Combining (\ref{traneig}) with (\ref{Lconj}) and Taylor expanding at
$u=0$, we find the momentum \be\label{momentum}
e^{2iP}=\frac{1}{2^{2L}}\,\Lambda(0)\bar{\Lambda}(0)=\prod_{j=1}^{M_u}\frac{u_j+i/2}{u_j-i/2}\prod_{j=1}^{M_v}\frac{v_j+i/2}{v_j-i/2}
\ee and energy, corresponding to the anomalous dimension $\gamma$,
\be\label{energy} E=\gamma=\lambda
^2\,\left(3L+\frac{d}{du}\ln(\Lambda(u)\bar{\Lambda}(u))\Big|_{u=0}\right)= \lambda
^2\left(\sum_{j=1}^{M_u}\frac{1}{u_j^2+\frac{1}{4}}+\sum_{j=1}^{M_v}\frac{1}{v_j^2+\frac{1}{4}}\right)\,.
\ee The state with $(K_u,K_r,K_v)$ Bethe roots belongs to the
$SU(4)$ representation with the Dynkin labels
$[L-2K_u+K_r,K_u+K_v-2K_r,L-2K_v+K_r]$. Consequently, the excitation
numbers must satisfy
\begin{equation}\label{kconstraints}
2K_u\leq L+K_r,\qquad 2K_v\leq L+K_r,\qquad 2K_r\leq K_u+K_v.
\end{equation}

In (\ref{momentum}) we see that the momentum carrying roots are the
outer roots, which contrasts with the $SU(4)$ spin chain found in
$\NN=4$ SYM which has the middle roots carrying the momentum
\cite{Minahan:2002ve}.  We can make a few simple checks on the
validity of the ansatz. First, we worked out the full set of
solutions for $L=2$ in appendix~\ref{dim2} and showed that it
matches with the spectrum of the Hamiltonian. There are also some
subsectors which can be easily identified for the spin chain and in
the Bethe ansatz equations. Let us choose a ground state operator
\be\label{gso} \tr[(Y^1Y^\dag_4)^L] \ee which is clearly symmetric
and traceless in the $Y^A$ and $Y^\dag_A$.  There are a few
subsectors of $SU(4)$ that are not mixed by anomalous dimension
matrix.  First there is an $SU(2)\times SU(2)$ subsector where the
scalars are $Y^1$ or $Y^2$ and the adjoints are $Y^\dag_3$ or
$Y^\dag_4$.  For this sector $K_{i,i+1}$ is always zero, so we are
left with two decoupled $SU(2)$ chains on the even and odd sites.
This corresponds to the absence of middle roots in the Bethe
equations, and in this case the Bethe equations reduce to two
decoupled Heisenberg spin chains with $L$ sites each.  The trace
condition couples the two chains by enforcing $e^{i(P_1+P_2)}=1$,
where $P_1$ and $P_2$ are the momentum for the first and second
Heisenberg chain.  There is also an $SU(3)$ sector where the scalars
are $Y^1$, $Y^2$ or $Y^3$, but the conjugates remain as $Y^\dag_4$.
In this case the even sites are a trivial chain and the odd sites
are part of an integrable $SU(3)$ chain.  This corresponds to the
absence of one of the outer roots and the Bethe equations can be
easily shown to reduce to that of an $SU(3)$ spin chain.  We could
also construct a different $SU(3)$ chain for the conjugate fields.

Since there are two types of roots that carry momentum, identifying
the elementary magnons is a little problematic.  To get some idea
of what to expect, let us again start with the ground state
operator in (\ref{gso}). On the string side, this should correspond
to a point-like string on $R\times CP^3$ located at
$(Z^1,Z^2,Z^3,Z^4)=(e^{i\om t},0,0,e^{-i\om t})$ where $\om$ is
proportional to the energy and $t$ is the world-sheet time which
can be gauge fixed to the target space time.  It is natural to
expect the elementary magnons to be associated with excitations
transverse to the motion.  Four of the transverse directions
correspond to rotations of $Z^1$ into $Z^2$ or $Z^3$, {\it or}
$\bar Z_4$ into $\bar Z_2$ or $\bar Z_3$.  In terms of the scalar
fields, this takes one of the $Y^1$ into a $Y^2$ or $Y^3$ {\it or}
one of the $Y_4^\dag$ to a $Y^\dag_2$ or $Y^\dag_3$.   If we choose
the simple root vectors as $\vec\al_1=(1,-1,0,0)$,
$\vec\al_2=(0,1,-1,0)$ and $\vec\al_3=(0,0,1,-1)$, then this
corresponds to subtracting off $\vec\al_1$, $\vec\al_1+\vec\al_2$,
$\vec \al_3+\vec\al_2$ or $\vec\al_3$ from the weights.  Hence
these elementary magnons are either a  momentum carrying root or a
momentum carrying root plus one middle root.  The last transverse
direction in $CP^3$ is a rotation of $Z^1$ into $Z^4$ and $\bar
Z_4$ into $\bar Z_1$.  On the gauge theory side this turns  a $Y^1$
into a $Y^4$ {\it and} a $Y^\dag_4$ into a $Y^\dag_1$.  The
combination of roots that give this is
$2\vec\al_1+2\vec\al_2+2\vec\al_3$.  But this actually produces two
charged zero pairs, so there is a smaller excitation with  half
this much.  Hence the last magnon has one each of the three types
of roots.  Since two of these roots carry momentum, one should
think of this magnon as a bound pair  of two of the other four
magnons.

\section{Extension to $OSp(2,2|6)$}

In this section we consider the extension of the $SU(4)$ spin chain
to the full superconformal group,
$OSp(2,2|6)$.  The extension of the Bethe equations is analogous to the $\NN=4$ SYM${}_4$ case in \cite{Beisert:2003yb}.  The bosonic subgroup is $SO(2,3)\times SO(6)$, the
product group of the three dimensional conformal group and the
$R$-symmetry group.  The fermionic elements are the 12 supersymmetry
generators $Q^{[AB]}_\al$ and the 12 superconformal generators
$S^{[AB]}_\al$, where the $A$ and $B$ indices are anti-symmetrized.

Following Kac's classification, this is a D(2,3) algebra with  a
nonunique Cartan matrix \be \left(\begin{array}{ccccc}-2&+1&&&\\
+1&&-1&&\\ &-1&+2&-1&-1\\ &&-1&+2&\\ &&-1&&+2
\end{array}\right)
\ee If we assume that the same bosonic roots carry the momentum then
the Dynkin diagram, including the Dynkin labels, is that in figure
6.  Note that one of the simple roots is fermionic and has invariant
length 0.
\begin{figure}[t]\label{ospfig}
\centerline{\includegraphics[totalheight=0.125\textheight]{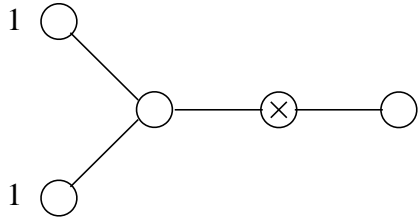}}
\caption{\label{ospdyn}\small One choice for the  $OSp(2,2|6)$
Dynkin diagram.  The Dynkin labels are taken from the $SU(4)$ spin
chain.}
\end{figure}

Given this diagram and Cartan matrix and following the general
recipe of \cite{Ogievetsky:1986hu}, the Bethe ansatz for the
$OSp(2,2|6)$ superalgebra is
\be\label{betheeqscf}
\left(\frac{u_j+i/2}{u_j-i/2}\right)^L&=&\prod_{k=1,k\ne j}^{M_u}\frac{u_j-u_k+i}{u_j-u_k-i} \prod_{k=1}^{M_r}\frac{u_j-r_k-i/2}{u_j-r_k+i/2}\nn\\
\left(\frac{v_j+i/2}{v_j-i/2}\right)^L&=&\prod_{k=1,k\ne
j}^{M_v}\frac{v_j-v_k+i}{v_j-v_k-i}
\prod_{k=1}^{M_r}\frac{v_j-r_k-i/2}{v_j-r_k+i/2}\nn\\
1&=& \prod_{k=1,k\ne j}^{M_r}\frac{r_j-r_k+i}{r_j-r_k-i}\prod_{k=1}^{M_u}\frac{r_j-u_k-i/2}{r_j-u_k+i/2}\prod_{k=1}^{M_v}\frac{r_j-v_k-i/2}{r_j-v_k+i/2}\prod_{k=1}^{M_s}\frac{r_j-s_k-i/2}{r_j-s_k+i/2}\nn\\
1&=&\prod_{k=1}^{M_r}\frac{s_j-r_k-i/2}{s_j-r_k+i/2}\prod_{k=1}^{M_w}\frac{s_j-w_k+i/2}{s_j-w_k-i/2}\nn\\
1&=&\prod_{k=1,k\ne j}^{M_w}\frac{w_j-w_k-i}{w_j-w_k+i}\prod_{k=1}^{M_s}\frac{w_j-s_k+i/2}{w_j-s_k-i/2}\,.
\ee

 The five bosonic charges in $OSp(2,2|6)$ can be grouped as $(-D-S,-D+S;J_1,J_2,J_3)$, where $D$ is the bare dimension, $S$ is the spin and $J_i$ are the three commuting $R$-charges in $SO(6)$.    The  ground state operator in (\ref{gso}) has charges $(-L,-L;L,L,0)$ and the charges of the simple root vectors are
\be
&&\vec\al_1=(0,0;0,1,-1)\,,\qquad\vec\al_2=(0,0;1,-1,0)\,,\qquad\vec\al_3=(0,0;0,1,1)\,\nn\\
&&\vec\beta=(1,-1;0,0,0)\,\qquad\vec\gamma=(0,1;-1,0,0)\,, \ee where
$\gamma$ is a fermionic root and the signature is $(--+++)$.
 The elementary magnons are the four discussed in the previous section as well as four fermionic magnons.  These last four have one momentum carrying root, either a $u$ or a $v$, as well as an $r$ and an $s$ root.  In addition the magnon may also include one $w$ root.  Hence an elementary fermionic magnon increases $D$ by $1/2$, increases or decreases $S$ by $1/2$, decreases $J_2$ by $1$ and increases or decreases $J_3$ by $1$.
 A covariant derivative does not correspond to an elementary magnon; instead this is a bound state of two fermionic roots.   All such bound states contain one $u$ and $v$  root, two $r$ and $s$ roots, and either zero, one or two roots, corresponding to a spin of $-1$, $0$ or $+1$.  One can also see this another way:  Unlike the $\NN=4$ SYM${}_4$ case, the $SL(2)$ sector in the superconformal Chern-Simons is not a closed sector.  In particular the combination $D_\mu Y^\dag_AY^B$ can mix into $\bar\psi^B\gamma_\mu\psi_A$, explicitly showing the two fermionic excitations.

Since the super Lie algebra has fermionic roots,  the Dynkin diagram
in figure 6 is not  the only choice we can make. A different grading
of roots can be found by grouping the charges as
$(J_1;-D-S,-D+S;J_2,J_3)$ and choosing the simple roots as \be
&&\vec\al_1=(0;0,0;1,-1)\,,\qquad\vec\al_2=(0;0,0;1,1)\,\nn\\
&&\vec\gamma_1=(0;0,1;-1,0)\,,\qquad
\vec\beta=(0;1,-1;0,0)\,\qquad\vec\gamma_2=(1;-1,0;0,0) \,. \ee Now
the super Dynkin diagram is the one in figure 7
\begin{figure}[t]\label{ospfig2}
\centerline{\includegraphics[totalheight=0.125\textheight]{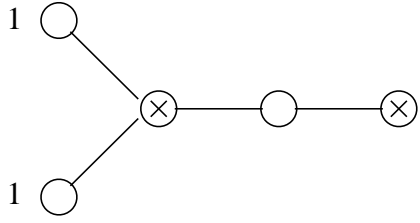}}
\caption{\label{ospdyn2}\small A different choice for the
$OSp(2,2|6)$ Dynkin diagram with two fermionic roots.}
\end{figure}
and the new Bethe equations are \be\label{betheeqscfalt}
\left(\frac{u_j+i/2}{u_j-i/2}\right)^L&=&\prod_{k=1,k\ne j}^{M_u}\frac{u_j-u_k+i}{u_j-u_k-i} \prod_{k=1}^{M_r}\frac{u_j-r_k-i/2}{u_j-r_k+i/2}\nn\\
\left(\frac{v_j+i/2}{v_j-i/2}\right)^L&=&\prod_{k=1,k\ne
j}^{M_v}\frac{v_j-v_k+i}{v_j-v_k-i}
\prod_{k=1}^{M_r}\frac{v_j-r_k-i/2}{v_j-r_k+i/2}\nn\\
1&=&\prod_{k=1}^{M_u}\frac{r_j-u_k-i/2}{r_j-u_k+i/2}\prod_{k=1}^{M_v}\frac{r_j-v_k-i/2}{r_j-v_k+i/2}\prod_{k=1}^{M_s}\frac{r_j-s_k+i/2}{r_j-s_k-i/2}\nn\\
1&=& \prod_{k=1,k\ne j}^{M_s}\frac{s_j-s_k-i}{s_j-s_k+i}\prod_{k=1}^{M_r}\frac{s_j-r_k+i/2}{s_j-r_k-i/2}\prod_{k=1}^{M_w}\frac{s_j-w_k+i/2}{s_j-w_k-i/2}\nn\\
1&=&\prod_{k=1}^{M_s}\frac{w_j-s_k+i/2}{w_j-s_k-i/2}\,, \ee where
$r$ and $w$ are now fermionic roots.  Of course, this system must be
equivalent to the one in (\ref{betheeqscf}), which can be shown
using the duality transformations in \cite{Tsuboi:1998ne} (see also
\cite{Beisert:2005di,Kazakov:2007fy}).  It is possible that this
choice of basis is more amenable to higher loop generalizations.
Figure \ref{ospdyn3} shows other bases for the simple roots, where
the Bethe equations can all be connected through duality
transformations.  The duality transformation \cite{Tsuboi:1998ne,
Beisert:2005di,Kazakov:2007fy} on the middle node produces a double link
between the momentum-carrying nodes in \ref{ospdyn3}b, which are
non-interacting in the original Dynkin diagram\footnote{ We  thank N.~Beisert for pointing this out to us.}.   The last
diagram in \ref{ospdyn3}c is found by dualizing one of the
momentum~carrying nodes in \ref{ospdyn3}b.  The two weights over the
left node signifies that both weights appear in the Bethe equations:
\begin{eqnarray}\label{}
\left(\frac{u_j-i/2}{u_j+i/2}\,\,\frac{u_j-3i/2}{u_j+3i/2}\right)^L
&=&
\prod_{k=1,k\neq j}^{M_u}\frac{u_j-u_k+2i}{u_j-u_k-2i}\,
\prod_{k=1}^{M_v}\frac{u_j-v_k-i}{u_j-v_k+i}\nonumber \\
\left(\frac{v_j+i/2}{v_j-i/2}\right)^L
&=&
\prod_{k=1}^{M_u}\frac{v_j-u_k-i}{v_j-u_k+i}\,
\prod_{k=1}^{M_r}\frac{v_j-r_k+i/2}{v_j-r_k-i/2}\nonumber \\
1
&=&
\prod_{k=1,k\neq j}^{M_r}\frac{r_j-r_k-i}{r_j-r_k+i}\,
\prod_{k=1}^{M_u}\frac{r_j-v_k+i/2}{r_j-v_k-i/2}\,
\prod_{k=1}^{M_s}\frac{r_j-s_k+i/2}{r_j-s_k-i/2}\nonumber \\
1
&=&
\prod_{k=1,k\neq j}^{M_s}\frac{s_j-s_k-i}{s_j-s_k+i}\,
\prod_{k=1}^{M_r}\frac{s_j-r_k+i/2}{s_j-r_k-i/2}\,
\prod_{k=1}^{M_w}\frac{s_j-w_k+i/2}{s_j-w_k-i/2}\nonumber \\
1
&=&
\prod_{k=1}^{M_s}\frac{w_j-s_k+i/2}{w_j-s_k-i/2}\,.
\end{eqnarray}
\begin{figure}[t]\label{ospfig3}
\centerline{\includegraphics[totalheight=0.12\textheight]{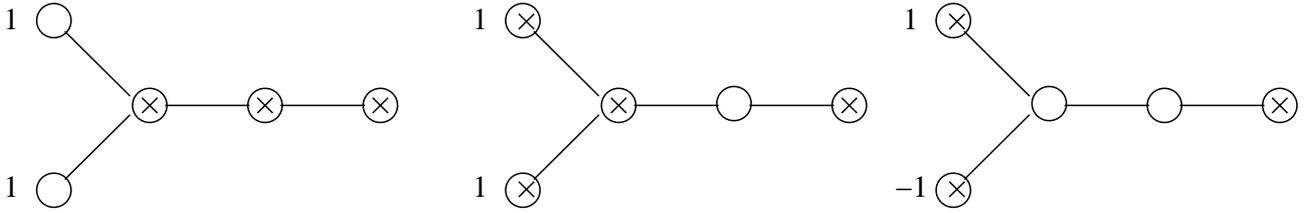}}
\caption{\label{ospdyn3}\small  Other $OSp(2,2|6)$ Dynkin diagrams.}
\end{figure}
The anomalous dimension for this choice of diagram is
\be\label{energy} \gamma= \lambda
^2\left(2L+\sum_{j=1}^{M_u}\left[\frac{1}{u_j^2+\frac{1}{4}}+\frac{3}{u_j^2+\frac{9}{4}}\right]-\sum_{j=1}^{M_v}\frac{1}{v_j^2+\frac{1}{4}}\right)\,.
\ee

We also note that  $OSP(2,2|6)$ has an $SU(2|3)$ subgroup with a diagram like\\
\centerline{\parbox{7cm}{\includegraphics[totalheight=0.07\textheight]{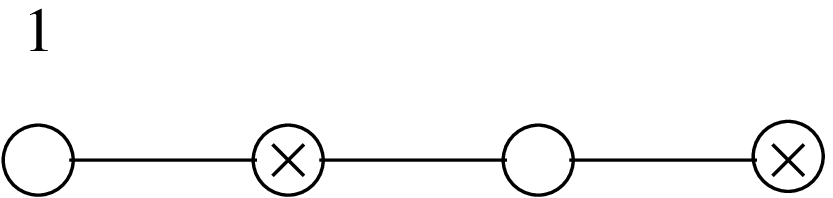}}}

\bigskip
\noindent This is the same diagram one finds for the $SU(2|3)$
subgroup of $SU(2,2|4)$ in $\NN=4$ SYM${}_4$ \cite{Beisert:2003ys}.
For higher loop calculations, one might expect to have the same set
of Bethe equations in this sector as the $\NN=4$ case
\cite{Beisert:2005fw}, but with $\la$ replaced by $\la^2$.  However,
the  dressing factors in \cite{Beisert:2006ib,Beisert:2006ez} might
need to be modified since the string action still contains an
overall factor of $\sqrt{\la}$.

\section{Summary and discussion}

We have shown that the ABJM $\mathcal{N}=6$ super-Chern-Simons  theory is
integrable at two loops, the lowest nontrivial order.  We also derived a set of Bethe equations for the
spectrum of two-loop anomalous dimensions. In conjunction with
classical integrability of the sigma-model on $AdS_4\times CP^3$,
the two loop integrability gives strong indications that the model
is integrable at any coupling. It might then be solvable in the
large-$N$ limit using an all-orders Bethe ansatz. We believe that one can extend our
results to higher loop orders along the lines of
\cite{Beisert:2003tq}, and perhaps to construct the asymptotic Bethe
ansatz equations at the non-perturbative level, as was been done
for $\mathcal{N}=4$ super-Yang-Mills in four dimensions
\cite{Beisert:2005fw,Beisert:2006ez,Beisert:2006ib}.

Even though we see no apparent relationship between
$\mathcal{N}=6$ super-Chern-Simons and $\mathcal{N}=4$ SYM$_4$, the
$AdS_4/CFT_3$ correspondence seems to be another instance where integrability plays an important role in the gauge/string duality .

\bigskip
\subsection*{Acknowledgments}
\bigskip
We would like to thank N.~Beisert, A.~Hanany, V.~Kazakov, C.~Kounnas, N.~Gromov, S.-J. Rey,
A. Torrielli, and P. Vieira for discussions. The work of K.Z. was supported in
part by the Swedish Research Council under contract 621-2007-4177,
and in part by the RFFI grant 06-02-17383 and the grant for support
of scientific schools NSh-3036.2008.2. The work of J.A.M. was
supported in part by the Swedish Research Council under contract
2006-3373 and the STINT foundation. J.A.M. thanks the CTP at MIT and
the Galileo Galilei Institute   for
Theoretical Physics 
for hospitality and INFN for partial support during the course of this work.


\appendix

\section{Contributions from  fermion loops and gauge\\ terms in Feynman
diagrams}\label{compferm}

In this appendix we compute the contribution of fermion loops  and  gauge contributions  to the
spin chain Hamiltonian.  The manifestly $SU(4)$ invariant fermion
couplings in the Lagrangian were computed in \cite{Benna:2008zy} and
are of the form 
\be\label{fermL}
\Lagr_{YY\psi\psi}&=&-\,\frac{i}{2}\tr \Big[Y_A^\dag Y^A\bar\psi^B\psi_B- \bar\psi^{B}Y^AY_A^\dag\psi_B+2\bar\psi^{B}Y^AY^\dag_B\psi_A-2Y^\dag_BY^A\bar\psi^{B}\psi_A\nn\\
&&\qquad\qquad+\eps^{ABCD}Y^\dag_A\psi_B^T\gamma_0Y^\dag_B\psi_D-\eps_{ABCD}Y^A\bar\psi^BY^C\psi^{D*}\Big]\,.
\ee The two loop planar graphs with a fermion loop are shown in
figure \ref{diagrams}b and \ref{diagrams}c.  Both diagrams can lead
to nontrivial interactions between the neighboring sites since
$SU(4)$ flavor is carried by the fermions.  However, only the graph
in \ref{diagrams}b has a log divergence. The only possible
interaction term is a contraction piece $K_{i,i+1}$ between
neighboring sites, and only the the third term in the first line of
(\ref{fermL}) can contribute to it.  The fourth term contributes to
the conjugate diagram.

Concentrating on just the contraction piece, we find the following
contribution to the operator renormalization  between scalars $i$ and $i+1$ coming from the counter term
\be
Z^{f}_{i,i+1}=-(-1)\left(-i\,\frac{4\pi}{k}\right)^2N^2\int
\frac{d^3p}{(2\pi)^3}\frac{d^3
q}{(2\pi)^3}\mbox{Tr}\left[\frac{i}{\psl+i\eps}\,\frac{i}{\qsl+i\eps}\right]\left(\frac{i}{p-q+i\eps}\right)^2
K_{i,i+1}\,.\nn\\
\ee
The trace refers to the fermion trace for three
dimensional Dirac fermions and the factor of $(-1)$ is for the
fermion loop.    After a Wick rotation and writing\, $2\, p\cdot
q=p^2+q^2-(p-q)^2$, we arrive at \be
Z^{f}_{i,i+1}=\left(4\pi\la\right)^2\int
\frac{d^3p}{(2\pi)^3}\frac{d^3 q}{(2\pi)^3}
\left[\frac{1}{p^2q^2(p-q)^2}-\frac{2}{p^2((p-q)^2)^2}\right]K_{i,i+1}\,.
\ee The second term inside the brackets does not contribute to the
anomalous dimension so we drop it. Dimensionally regulating the
integral, inserting a small mass term $\mu$ to act as an infrared
cutoff,  and inserting a Feynman parameterization we get \be
Z^{f}_{i,i+1}&=&\left(4\pi\la\right)^2\frac{1}{64\pi^3}\int_0^\infty
\frac{d\rho}{\rho^{1-\veps}}e^{-\rho\mu^2}\int_0^1 dx\int_0^x dy
[x(1-x)+y(1-y)-xy]^{-3/2}K_{i,i+1}\nn\\
&=&\frac{1}{2}\la^2\Gamma(\veps)\mu^{-2\veps}K_{i,i+1}\,. \ee Since
$\veps^{-1}\sim \ln \Lambda^2$, this contribution to the anomalous
dimension from all  neighboring sites  is \be
\Gamma_{f}=\sum_{i=1}^{2L}\frac{d}{d\ln\Lambda}Z^{f}_{i,i+1}=\la^2\sum_{i}^{2L}K_{i,i+1}\,.
\ee

Diagrams containing gauge boson propagators are shown in figures \ref{diagrams}d, \ref{diagrams}e, \ref{diagrams}f and \ref{diagrams}g, but only \ref{diagrams}d will contribute to the anomalous dimension.  The gauge propagators are given by
\be
\frac{2\pi}{k}\frac{p_\mu\eps^{\mu\nu\s}}{p^2}\,,
\ee
and only one of the $SU(N)$ gauge bosons will contribute to the planar diagram (the other $SU(N)$ contributes to the conjugate diagram.)
Hence the conribution to the operator normalization between scalars $i$ and $i+1$ from the diagram in \ref{diagrams}d is
\be
Z^{g}_{i,i+1}&=&-(+i)\left(\frac{2\pi}{k}\right)^2N^2\int
\frac{d^3p}{(2\pi)^3}\frac{d^3
q}{(2\pi)^3}(2iq_\nu+ip_\nu)(2iq_{\nu'}+ip_{\nu'})\frac{p_\mu{\eps^{\mu\nu}}_\s}{p^2}\frac{p_{\mu'}\eps^{\mu'\s\nu'}}{p^2}\nn\\
&&\qquad\qquad\qquad\qquad\qquad\qquad\times\left(\frac{i}{p^2}\right)^2\frac{i}{(p+q)^2}
K_{i,i+1}\,, \ee
where the factor of $(+i)$ comes from the four-point vertex.  This then gives
\be
Z^{g}_{i,i+1}=-\left(4\pi\la\right)^2\int
\frac{d^3p}{(2\pi)^3}\frac{d^3 q}{(2\pi)^3}\,
\frac{p^2q^2-(p\cdot q)^2}{p^4q^4(p+q)^2}\,K_{i,i+1}\,.
\ee
If we write
\be
p^2q^2-(p\cdot q)^2=\half p^2q^2+\half(p+q)^2(p^2+q^2)-\sfrac{1}{4}(p^4+q^4+(p+q)^4)\,,
\ee
only the first term will contribute to the log.  Following the arguments for the fermion loop, we can quickly see that
\be
Z^{g}_{i,i+1}&=&-\frac{1}{4}\la^2\Gamma(\veps)\mu^{-2\veps}K_{i,i+1} \ee
and so this contribution to the anomalous dimension is
\be
\Gamma_{g}=\sum_{i=1}^{2L}\frac{d}{d\ln\Lambda}Z^{g}_{i,i+1}=-\frac{1}{2}\,\la^2\sum_{i}^{2L}K_{i,i+1}\,.
\ee

The diagram in \ref{diagrams}g is nonzero, but only has a linear divergence and no log divergence.  The diagrams in \ref{diagrams}e and \ref{diagrams}f are zero because the momentum in the top scalar line is the same as the gauge momentum, and so they both have  $\eps^{\mu\nu\s}p_\mu p_\nu$ factors.

Combining $\Gamma_f$ and $\Gamma_g$, we get
\be
\Gamma_f+\Gamma_g=\frac{1}{2}\,\la^2\sum_{i}^{2L}K_{i,i+1}\,,
\ee
precisely canceling  the nearest neighbor term from the six-point graph.

\section{Chiral primaries and spherical functions on
$CP^3$}\label{appspfun}

Any chiral primary operator, (\ref{ops}) with symmetric traceless
$\chi _{A_1\ldots A_L}^{B_1\ldots B_L}$, defines a function on
$CP^3$:
\begin{equation}\label{spfun}
\chi (z,\bar{z})=\chi _{A_1\ldots A_L}^{B_1\ldots B_L}z^{A_1}\ldots
z^{A_L}\bar{z}_{B_1}\ldots \bar{z}_{B_L},
\end{equation}
where $z,\bar{z}$ are homogeneous coordinates constrained by
$z^A\bar{z}_A=1$, $z^A\sim \,{\rm e}\,^{i\varphi }z^A$,
$\bar{z}_A\sim \,{\rm e}\,^{-i\varphi }\bar{z}_A$. The
Laplace-Beltrami operator on $CP^3$ is the $U(3)$ Casimir. In terms
of the $U(3)$ generators,
\begin{equation}\label{}
L^A_B=z^A\,\frac{\partial }{\partial z^B}-\bar{z}_B\,\frac{\partial }{\partial
\bar{z}_A}\,,
\end{equation}
the Laplacian is
\begin{equation}\label{}
\Delta =\frac{1}{2}\,L^A_BL^B_A.
\end{equation}
It is easy to check that the function (\ref{spfun}) is its
eigenstate:
\begin{equation}\label{}
\Delta\,\chi (z,\bar{z})=L(L+3)\chi (z,\bar{z}).
\end{equation}

\section{Dimension-two operators}\label{dim2}

In this appendix we explicitly diagonalize the Hamiltonian (\ref{})
for the spin chain of length $4$, first by brute force, and then
with the help of the Bethe ansatz equations. For the sake of
generality we temporarily relax the trace condition (\ref{}). We
will indicate which operators satisfy it but will compute the whole
spectrum, including the states with non-zero total
momentum\footnote{For length four, the shift operator $\,{\rm
e}\,^{2iP}$ can have only two eigenvalues: $+1$ or $-1$.}.

The length-four Hilbert space decomposes as
$\mathbf{4}\otimes\bar{\mathbf{4}}\otimes\mathbf{4}\otimes\bar{\mathbf{4}}
=\mathbf{1}^2\oplus\mathbf{15}^4\oplus\mathbf{20}\oplus\mathbf{45}\oplus
\bar{\mathbf{45}}\oplus\mathbf{84}$. The $\mathbf{84}$ is the chiral
primary with totally symmetric traceless wavefunction and zero
energy:
\begin{equation}\label{}
\mathbf{84}:~\chi _{(AB)}^{(CD)}-{\rm traces},~\gamma_\mathbf{84} =0,~\,{\rm e}\,^{2iP}=1.
\end{equation}
The $\mathbf{45}$ and $\bar{\mathbf{45}}$ are symmetric in one pair
of indices and anti-symmetric in the other. They do not correspond
to any operators because of the trace condition. The permutation
operator centered at the odd/even sites now yields a $-1$ and
doubles the constant term in the Hamiltonian:
\begin{eqnarray}\label{e45}
&&\mathbf{45}:~\chi ^{(AB)}_{[CD]}-{\rm traces},~\gamma_\mathbf{45} =4\lambda^2
,~\,{\rm e}\,^{2iP}=-1,\nonumber \\
&&\bar{\mathbf{45}}:~\chi ^{[AB]}_{(CD)}-{\rm traces},~\gamma_{\bar{\mathbf{45}}} =4\lambda^2
,~\,{\rm e}\,^{2iP}=-1.
\end{eqnarray}
The
$\mathbf{20}$ is anti-symmetric in each pair of indices and the
constant term is now doubled on all the sites:
\begin{equation}\label{e20}
\mathbf{20}:~\chi _{[CD]}^{[AB]}-{\rm traces},~\gamma
_\mathbf{20}=8\lambda^2 ,\,{\rm e}\,^{2iP}=1.
\end{equation}

The non-trivial mixing first occurs in the adjoint representation,
the $\mathbf{15}$. Let us denote the four adjoint states by
\begin{eqnarray}\label{}
&&\left|1\right\rangle_\mathbf{15}:~\chi ^{CA}_{CB}-{\rm trace},\nonumber \\
&&\left|2\right\rangle_\mathbf{15}:~\chi ^{AC}_{CB}-{\rm trace},\nonumber \\
&&\left|3\right\rangle_\mathbf{15}:~\chi ^{AC}_{BC}-{\rm trace},\nonumber \\
&&\left|4\right\rangle_\mathbf{15}:~\chi ^{CA}_{BC}-{\rm trace}.\nonumber
\end{eqnarray}
The Hamiltonian and momentum act in this basis as
\begin{eqnarray}\label{}
\Gamma \left|n\right\rangle&=&5\lambda^2 \left|n\right\rangle
+\lambda^2 \left|n+2\right\rangle\nonumber \\
\,{\rm e}\,^{2iP}\left|n\right\rangle
&=&\left|n+2\right\rangle.
\end{eqnarray}
The eigenstates $\left|1\right\rangle\pm\left|3\right\rangle$ and
$\left|2\right\rangle\pm\left|4\right\rangle$ are doubly degenerate
with the eigenvalues
\begin{equation}\label{e15}
\mathbf{15}:
~\gamma _\mathbf{15}^{(\pm)}=(5\pm 1)\lambda^2,~\,{\rm
e}\,^{2iP}=\pm 1.
\end{equation}

The two singlets,
\begin{eqnarray}\label{}
&&\left|1\right\rangle_\mathbf{1}:~\chi ^{AB}_{AB},\nonumber \\
&&\left|2\right\rangle_\mathbf{1}:~\chi ^{AB}_{BA},\nonumber
\end{eqnarray}
both have zero total momentum and mix according to
\begin{equation}\label{}
\left.\Gamma \right|_\mathbf{1}=\lambda^2
\begin{pmatrix}
  6  & 4  \\
   4 & 6  \\
\end{pmatrix}.
\end{equation}
The eigenvalues are
\begin{eqnarray}\label{e1}
\mathbf{1}:&&\gamma _\mathbf{1}^{(1)}=2\lambda^2,~\,{\rm
e}\,^{2iP}=1,\nonumber \\
&&\gamma _\mathbf{1}^{(2)}=10\lambda^2 ,~\,{\rm
e}\,^{2iP}=1.
\end{eqnarray}
\begin{table}[t]
\begin{center}
\begin{tabular}{|c|c|}
\hline
 Anomalous dimension & $SU(4)$ representation
 \\
 \hline
 $0$ & $\mathbf{84}$ \\
 $2\lambda^2$ & $\mathbf{1}$ \\
 $6\lambda^2 $ & $\mathbf{15}$ \\
 $6\lambda^2 $ & $\mathbf{15}$ \\
 $8\lambda^2 $ & $\mathbf{20}$ \\
 $10\lambda^2 $ & $\mathbf{1}$ \\
 \hline
\end{tabular}
\end{center}
\caption{\small The spectrum of operators at $L=2$.}
\label{dregimes}
\end{table}
The spectrum of dimension two operators is summarized in
table~\ref{dregimes}.

Let us see how the Bethe equations (\ref{}) reproduce this spectrum.
The conditions (\ref{}) admit the following root configurations
$(K_u,K_r,K_v)$:
\begin{eqnarray}\label{}
\mathbf{84}:&&(0,0,0)\nonumber \\
\mathbf{45}:&&(1,0,0)\nonumber \\
\bar{\mathbf{45}}:&&(0,0,1)\nonumber \\
\mathbf{20}:&&(1,0,1)\nonumber \\
\mathbf{15}:&&(1,1,1)\nonumber \\
\mathbf{1}:&&(2,2,2).\nonumber
\end{eqnarray}
For the configurations with only one $u$ root or only one $v$ root
(the $\mathbf{45}$ and the $\bar{\mathbf{45}}$), the Bethe equations
admit a unique solution: $u_1=0$ or $v_1=0$, whose energy (\ref{})
is $\gamma _{\mathbf{45}/\bar{\mathbf{45}}}=4\lambda^2 $, in
agreement with (\ref{e45}). These states can be combined:
$u_1=0=v_1$, which yields the $\mathbf{20}$ with energy $\gamma
_\mathbf{20}=8\lambda^2 $.

The Bethe equations for the $\mathbf{15}$ with $u_1\equiv u$,
$r_1\equiv r$ and $v_1\equiv v$ are
\begin{equation}\label{}
\left(\frac{u+\frac{i}{2}}{u-\frac{i}{2}}\right)^2=\frac{u-r-\frac{i}{2}}{u-r+\frac{i}{2}}\,,
\qquad
1=\frac{r-u-\frac{i}{2}}{r-u+\frac{i}{2}}\,\,\frac{r-v-\frac{i}{2}}{r-v+\frac{i}{2}}\,,
\qquad
\left(\frac{v+\frac{i}{2}}{v-\frac{i}{2}}\right)^2=\frac{v-r-\frac{i}{2}}{v-r+\frac{i}{2}}\,.
\end{equation}
They have four solutions:
\begin{equation}\label{}
u=v=r=\pm\frac{1}{2}\,,~\gamma _\mathbf{15}^{(-)}=4\lambda^2 ;\qquad
u=-v=\pm\frac{1}{2\sqrt{3}}\,,~r=0,~\gamma
_\mathbf{15}^{(+)}=6\lambda^2,
\end{equation}
which matches with (\ref{e15}).

The situation with singlets is more complicated. There is a regular
solution with $u_1=-u_2\equiv u$, $r_1=-r_2\equiv r$, and
$v_1=-v_2=u$ that satisfy
\begin{equation}\label{}
\frac{u+\frac{i}{2}}{u-\frac{i}{2}}=\frac{u-r-\frac{i}{2}}{u-r+\frac{i}{2}}\,\,
\frac{u+r-\frac{i}{2}}{u+r+\frac{i}{2}}\,,\qquad
1=\frac{r+\frac{i}{2}}{r-\frac{i}{2}}\left(
\frac{r-u-\frac{i}{2}}{r-u+\frac{i}{2}}\,\,\frac{r+u-\frac{i}{2}}{r+u+\frac{i}{2}}
\right)^2.
\end{equation}
These equations have a unique solution:
\begin{equation}\label{}
u=\sqrt{\frac{3}{20}},~r=\frac{1}{\sqrt{5}}\,,~\gamma
_\mathbf{1}^{(2)}=10\lambda^2 .
\end{equation}
The other singlet corresponds to a singular distribution of roots
\cite{Beisert:2003xu,Beisert:2004hm}:
\begin{equation}\label{}
u_{1,2}=i\left(\pm\frac{1}{2}+\varepsilon \pm\,\delta\, \right)=v_{1,2},\qquad
r_1\equiv r=-r_2,
\end{equation}
which solves the Bethe equations in the limit $\varepsilon
\rightarrow 0$ with $\delta \ll\varepsilon $, when both sides of the
Bethe equations simultaneously turn to zero or to infinity. The
balance of infinities determines $\delta $ in terms of $\varepsilon
$:
\begin{equation}\label{}
\delta =\frac{r^2}{1+r^2}\,\varepsilon ^2.
\end{equation}
The middle-node equation is non-singular and gives:
\begin{equation}\label{}
r=\frac{i}{\sqrt{3}}\,.
\end{equation}
In the energy (\ref{}) the $1/\varepsilon $ singularity cancels. It
is important to keep the $O(\varepsilon ^2)$ terms to get the
finite part right:
\begin{equation}\label{}
\gamma _\mathbf{1}^{(1)}=2\lambda^2 \lim_{\varepsilon \rightarrow 0}
\left[
\frac{1}{\frac{1}{4}-\left(\frac{1}{2}+\varepsilon -\frac{\varepsilon ^2}{2}\right)^2}
+\frac{1}{\frac{1}{4}-\left(\frac{1}{2}-\varepsilon -\frac{\varepsilon ^2}{2}\right)^2}
\right]=2\lambda^2 ,
\end{equation}
which agrees with (\ref{e1}).

\bibliographystyle{nb}
\bibliography{refs}

\begin{thebibliography}{10}
\ifx\href\asklfhas\newcommand{\href}[2]{#2}\fi
\raggedright
\small
\parskip 0pt

\bibitem{Aharony:2008ug}
O.~Aharony, O.~Bergman, D.~L.~Jafferis and J.~Maldacena,
\textit{``{N=6 superconformal Chern-Simons-matter theories, M2-branes and their
  gravity duals}''},
\href{http://arXiv.org/abs/0806.1218}{\texttt{0806.1218}}.
%
\bibitem{Schwarz:2004yj}
J.~H.~Schwarz,
\textit{``{Superconformal Chern-Simons theories}''},
\textsf{JHEP~0411,~078~(2004)},
\href{http://arXiv.org/abs/hep-th/0411077}{\texttt{hep-th/0411077}}.
%
\bibitem{Bagger:2006sk}
J.~Bagger and N.~Lambert,
\textit{``{Modeling multiple M2's}''},
\textsf{Phys.~Rev.~D75,~045020~(2007)},
\href{http://arXiv.org/abs/hep-th/0611108}{\texttt{hep-th/0611108}}.
%
\bibitem{Gustavsson:2007vu}
A.~Gustavsson,
\textit{``{Algebraic structures on parallel M2-branes}''},
\href{http://arXiv.org/abs/0709.1260}{\texttt{0709.1260}}.
%
\bibitem{Bagger:2007jr}
J.~Bagger and N.~Lambert,
\textit{``{Gauge Symmetry and Supersymmetry of Multiple M2-Branes}''},
\textsf{Phys.~Rev.~D77,~065008~(2008)},
\href{http://arXiv.org/abs/0711.0955}{\texttt{0711.0955}}.
%
\bibitem{Bagger:2007vi}
J.~Bagger and N.~Lambert,
\textit{``{Comments On Multiple M2-branes}''},
\textsf{JHEP~0802,~105~(2008)},
\href{http://arXiv.org/abs/0712.3738}{\texttt{0712.3738}}.
%
\bibitem{VanRaamsdonk:2008ft}
M.~Van~Raamsdonk,
\textit{``{Comments on the Bagger-Lambert theory and multiple M2- branes}''},
\textsf{JHEP~0805,~105~(2008)},
\href{http://arXiv.org/abs/0803.3803}{\texttt{0803.3803}}.
%
\bibitem{Distler:2008mk}
J.~Distler, S.~Mukhi, C.~Papageorgakis and M.~Van~Raamsdonk,
\textit{``{M2-branes on M-folds}''},
\textsf{JHEP~0805,~038~(2008)},
\href{http://arXiv.org/abs/0804.1256}{\texttt{0804.1256}}.
%
\bibitem{Ho:2008ei}
P.-M.~Ho, Y.~Imamura and Y.~Matsuo,
\textit{``{M2 to D2 revisited}''},
\textsf{JHEP~0807,~003~(2008)},
\href{http://arXiv.org/abs/0805.1202}{\texttt{0805.1202}}.
%
\bibitem{Gomis:2008be}
J.~Gomis, D.~Rodriguez-Gomez, M.~Van~Raamsdonk and H.~Verlinde,
\textit{``{Supersymmetric Yang-Mills Theory From Lorentzian Three-Algebras}''},
\href{http://arXiv.org/abs/0806.0738}{\texttt{0806.0738}}.
%
\bibitem{Benna:2008zy}
M.~Benna, I.~Klebanov, T.~Klose and M.~Smedback,
\textit{``{Superconformal Chern-Simons Theories and $AdS_4/CFT_3$
  Correspondence}''},
\textsf{JHEP~0809,~072~(2008)},
\href{http://arXiv.org/abs/0806.1519}{\texttt{0806.1519}}.
%
\bibitem{Nishioka:2008gz}
T.~Nishioka and T.~Takayanagi,
\textit{``{On Type IIA Penrose Limit and N=6 Chern-Simons Theories}''},
\textsf{JHEP~0808,~001~(2008)},
\href{http://arXiv.org/abs/0806.3391}{\texttt{0806.3391}}.
%
\bibitem{Astolfi:2006is}
D.~Astolfi, V.~Forini, G.~Grignani and G.~W.~Semenoff,
\textit{``{Finite size corrections and integrability of N = 2 SYM and DLCQ
  strings on a pp-wave}''},
\textsf{JHEP~0609,~056~(2006)},
\href{http://arXiv.org/abs/hep-th/0606193}{\texttt{hep-th/0606193}}.
%
\bibitem{Minahan:2002ve}
J.~A.~Minahan and K.~Zarembo,
\textit{``The bethe-ansatz for {$\mathcal{N}=\mathord{}$4} super yang-mills''},
\textsf{JHEP~0303,~013~(2003)},
\href{http://arXiv.org/abs/hep-th/0212208}{\texttt{hep-th/0212208}}.
%
\bibitem{Beisert:2003jj}
N.~Beisert,
\textit{``{The complete one-loop dilatation operator of N = 4 super Yang-Mills
  theory}''},
\textsf{Nucl.~Phys.~B676,~3~(2004)},
\href{http://arXiv.org/abs/hep-th/0307015}{\texttt{hep-th/0307015}}.
%
\bibitem{Beisert:2003yb}
N.~Beisert and M.~Staudacher,
\textit{``The {$\mathcal{N}=\mathord{}$4} sym integrable super spin chain''},
\textsf{Nucl.~Phys.~B670,~439~(2003)},
\href{http://arXiv.org/abs/hep-th/0307042}{\texttt{hep-th/0307042}}.
%
\bibitem{Beisert:2003tq}
N.~Beisert, C.~Kristjansen and M.~Staudacher,
\textit{``The dilatation operator of {$\mathcal{N}=\mathord{}$4} conformal
  super yang-mills theory''},
\textsf{Nucl.~Phys.~B664,~131~(2003)},
\href{http://arXiv.org/abs/hep-th/0303060}{\texttt{hep-th/0303060}}.
%
\bibitem{Beisert:2005fw}
N.~Beisert and M.~Staudacher,
\textit{``Long-range $psu(2,2|4)$ bethe ansaetze for gauge theory and
  strings''},
\textsf{Nucl.~Phys.~B727,~1~(2005)},
\href{http://arXiv.org/abs/hep-th/0504190}{\texttt{hep-th/0504190}}.
%
\bibitem{Beisert:2006ez}
N.~Beisert, B.~Eden and M.~Staudacher,
\textit{``{Transcendentality and crossing}''},
\textsf{J.~Stat.~Mech.~0701,~P021~(2007)},
\href{http://arXiv.org/abs/hep-th/0610251}{\texttt{hep-th/0610251}}.
%
\bibitem{Beisert:2006ib}
N.~Beisert, R.~Hernandez and E.~Lopez,
\textit{``A crossing-symmetric phase for $ads_5 \times s^5$ strings''},
\textsf{JHEP~0611,~070~(2006)},
\href{http://arXiv.org/abs/hep-th/0609044}{\texttt{hep-th/0609044}}.
%
\bibitem{Gaiotto:2007qi}
D.~Gaiotto and X.~Yin,
\textit{``{Notes on superconformal Chern-Simons-matter theories}''},
\textsf{JHEP~0708,~056~(2007)},
\href{http://arXiv.org/abs/0704.3740}{\texttt{0704.3740}}.
%
\bibitem{Faddeev:1985qu}
L.~D.~Faddeev and N.~Y.~Reshetikhin,
\textit{``{Integrability of the principal chiral field model in
  (1+1)-dimension}''},
\textsf{Ann.~Phys.~167,~227~(1986)}.
%
\bibitem{Destri:1987ug}
C.~Destri and H.~J.~de~Vega,
\textit{``{Light cone lattices and the exact solution of chiral fermion and
  sigma models}''},
\textsf{J.~Phys.~A22,~1329~(1989)}.
%
\bibitem{deVega:1991rc}
H.~J.~de~Vega and F.~Woynarovich,
\textit{``{New integrable quantum chains combining different kinds of
  spins}''},
\textsf{J.~Phys.~A25,~4499~(1992)}.
%
\bibitem{abadrios}
J.~Abad and M.~Rios,
\textit{``{Integrable $su(3)$ spin chain combining different
  representations}''},
\textsf{J.~Phys.~A30,~5887~(1997)},
\href{http://arXiv.org/abs/cond-mat/9706136}{\texttt{cond-mat/9706136}}.
%
\bibitem{Martins99}
M.~J.~Martins,
\textit{``{Integrable mixed vertex models from braid-monoid algebra}''},
\href{http://arXiv.org/abs/solv-int/9903006}{\texttt{solv-int/9903006}}.
%
\bibitem{Ribeiro:2005kn}
G.~A.~P.~Ribeiro and M.~J.~Martins,
\textit{``{Algebraic Bethe ansatz for an integrable $U_q[Sl(n|m)]$ vertex model
  with mixed representations}''},
\textsf{Nucl.~Phys.~B738,~391~(2006)},
\href{http://arXiv.org/abs/nlin/0512035}{\texttt{nlin/0512035}}.
%
\bibitem{Kulish:1983rd}
P.~P.~Kulish and N.~Y.~Reshetikhin,
\textit{``{Diagonalization of $GL(N)$ invariant transfer matrices and quantum
  $N$ wave system (Lee model)}''},
\textsf{J.~Phys.~A16,~L591~(1983)}.
%
\bibitem{Arnaudon:2004vd}
A.~Arnaudon, N.~Crampe, A.~Doikou, L.~Frappat and E.~Ragoucy,
\textit{``{Analytical Bethe Ansatz for closed and open $gl(n)$-spin chains in
  any representation}''},
\textsf{J.~Stat.~Mech.~0502,~P007~(2005)},
\href{http://arXiv.org/abs/math-ph/0411021}{\texttt{math-ph/0411021}}.
%
\bibitem{Faddeev:1996iy}
L.~D.~Faddeev,
\textit{``{How Algebraic Bethe Ansatz works for integrable model}''},
\href{http://arXiv.org/abs/hep-th/9605187}{\texttt{hep-th/9605187}}.
%
\bibitem{Ogievetsky:1986hu}
E.~Ogievetsky and P.~Wiegmann,
\textit{``{Factorized s matrix and the bethe ansatz for simple Lie groups}''},
\textsf{Phys.~Lett.~B168,~360~(1986)}.
%
\bibitem{Tsuboi:1998ne}
Z.~Tsuboi,
\textit{``{Analytic Bethe Ansatz And Functional Equations Associated With Any
  Simple Root Systems Of The Lie Superalgebra $sl(r+1|s+1)$}''},
\textsf{Physica~A252,~565~(1998)}.
%
\bibitem{Beisert:2005di}
N.~Beisert, V.~A.~Kazakov, K.~Sakai and K.~Zarembo,
\textit{``{Complete spectrum of long operators in N = 4 SYM at one loop}''},
\textsf{JHEP~0507,~030~(2005)},
\href{http://arXiv.org/abs/hep-th/0503200}{\texttt{hep-th/0503200}}.
%
\bibitem{Kazakov:2007fy}
V.~Kazakov, A.~Sorin and A.~Zabrodin,
\textit{``{Supersymmetric Bethe ansatz and Baxter equations from discrete
  Hirota dynamics}''},
\textsf{Nucl.~Phys.~B790,~345~(2008)},
\href{http://arXiv.org/abs/hep-th/0703147}{\texttt{hep-th/0703147}}.
%
\bibitem{Beisert:2003ys}
N.~Beisert,
\textit{``{The $su(2|3)$ dynamic spin chain}''},
\textsf{Nucl.~Phys.~B682,~487~(2004)},
\href{http://arXiv.org/abs/hep-th/0310252}{\texttt{hep-th/0310252}}.
%
\bibitem{Beisert:2003xu}
N.~Beisert, J.~A.~Minahan, M.~Staudacher and K.~Zarembo,
\textit{``{Stringing spins and spinning strings}''},
\textsf{JHEP~0309,~010~(2003)},
\href{http://arXiv.org/abs/hep-th/0306139}{\texttt{hep-th/0306139}}.
%
\bibitem{Beisert:2004hm}
N.~Beisert, V.~Dippel and M.~Staudacher,
\textit{``{A novel long range spin chain and planar N = 4 super Yang-
  Mills}''},
\textsf{JHEP~0407,~075~(2004)},
\href{http://arXiv.org/abs/hep-th/0405001}{\texttt{hep-th/0405001}}.
%
\end{thebibliography}

\end{document}